# 6 VALUE-BASED INVENTORY MANAGEMENT

Grzegorz MICHALSKI[*]

## Abstract

*The basic financial purpose of a firm is to maximize its value. An inventory management system should also contribute to realization of this basic aim. Many current asset management models currently found in financial management literature were constructed with the assumption of book profit maximization as basic aim. However these models could lack what relates to another aim, i.e., maximization of enterprise value. This article presents a modified value-based inventory management model.*

## 1. Introduction

The basic financial aim of an enterprise is maximization of its value. At the same time, a large both theoretical and practical meaning has the research for determinants increasing the firm value. Most financial literature contains information about numerous factors influencing the value. Among those factors is the net working capital and elements creating it, such as the level of cash tied in accounts receivable, inventories and operational cash balances. A large majority of classic financial models proposals, relating to the optimum current assets management, were constructed with net profit maximization in view. In order to make these models more suitable for firms, which want to maximize their value, some of them must be reconstructed. In the sphere of inventory management, the estimation of the influence of changes in a firm's decisions is a compromise between limiting risk by having greater inventory and limiting the costs of inventory. It is the essential problem of the corporate financial management.

The basic financial inventory management aim is holding the inventory to a minimally acceptable level in relation to its costs. Holding inventory means using capital to

[*] Ph.D. Grzegorz Michalski, Wroclaw University of Economics, Department of Corporate Finance and Value Management, ul. Komandorska 118/120, pok. 704-Z, PL53-345 Wroclaw, Poland, e-mail: Grzegorz.Michalski@ae.wroc.pl ; http://michalskig.com/

*Value-Based Inventory Management*

finance inventory and links with inventory storage, insurance, transport, obsolescence, wasting and spoilage costs. However, maintaining a low inventory level can, in turn, lead to other problems with regard to meeting supply demands.

## 2. VALUE-BASED INVENTORY MANAGEMENT

If advantages from holding inventory on a level defined by the firm will be greater than the negative influence of an opportunity costs from its holding, then the firm's value will grow. Change of the inventory level affects the firm value. To measure that value, we use a formula based on the assumption that the firm value is a sum of future free cash flows to firm (*FCFF*) discounted by cost of capital financing the firm:

$$\Delta V_p = \sum_{t=1}^{n} \frac{\Delta FCFF_t}{(1+k)^t}, \qquad (1)$$

Where: $\Delta V_p$ = firm value growth; $\Delta FCFF_t$ = future free cash flow growth in period *t*, and $k$ = discount rate[1].

Future free cash flow we have as:

$$FCFF_t = (CR_t - CE_t - NCE) \times (1-T) + NCE - Capex - \Delta NWC_t \qquad (2)$$

Where: $CR_t$ = cash revenues on sale; $CE_t$ = cash expenses resulting from fixed and variable costs in time *t*; *NCE* = non cash expenses; *T* = effective tax rate; $\Delta NWC$ = net working growth; and *Capex* = capital expenditure resulting from operational investments growth (money used by a firm to acquire or upgrade physical assets such as property, industrial buildings, or equipment).

The similar conclusions, about the results of the change inventory management policy on the firm value, can be estimated on the basis of an economic value added, informing about the size of the residual profit (the added value) enlarged the value of the firm in the period:

$$EVA = NOPAT - k \times (NWC + OI) \qquad (3)$$

Where: *EVA* = economic value added; *NWC* = net working capital; *OI* = long-term operating investments; and *NOPAT* = net operating profit after taxes, estimated on the basis of the formula:

$$NOPAT = (CR_t - CE_t - NCE) \times (1-T) \qquad (4)$$

The net working capital (NWC) is the part of current assets, financed with fixed capitals. The net working capital (current assets less current liabilities) results from lack of synchronization of the formal rising receipts and the real cash receipts from each sale. Net working capital also results from divergence during time of rising costs and time, from the real outflow of cash when a firm pays its accounts payable.

$$NWC = CA - CL = AAR + INV + G - AAP \qquad (5)$$

---

[1] *To estimate changes in accounts receivable levels, we accept a discount rate equal to the average weighted cost of capital (WACC). Such changes and their results are strategic and long term in their character, although they refer to accounts receivable and short run area decisions (T.S. Maness 1998, s. 62-63).*

Where: *NWC* = net working capital; *CA* = current assets; *CL* = current liabilities; *AAR* = average level of accounts receivables; *INV* = inventory; *G* = cash and cash equivalents; and *AAP* = average level of accounts payables.

During estimation of the free cash flows the holding and increasing of net working capital ties money used for financing it. If net working capital increase, the firm must tie much money and it decrease free cash flows. The production level growth usually causes the necessity of an enlargement of cash levels, inventories, and accounts receivable. Part of this growth will be covered with current liabilities. For current liabilities also usually automatically grow up together with the growth of production. The rest (which is noted as net working capital growth) will require other forms of financing.

The inventory management policy decisions, create the new inventory level in a firm. It has the influence on the firm value. It is the result of opportunity costs of money tied in with inventory and generally of costs of inventory managing. Both the first and the second involve modification of future free cash flows, and in consequence the firm value changes. On Figure 1, we have the influence of inventory management decisions on the firm value. These decisions change the future free cash flows (*FCFF*). These decisions could also have influence on the life of the firm (*t*) (by the operational risk, which is the result of the possibility to break production cycles if the inventory level is too low), and rate of the cost of capital financing the firm (*k*). The changes of these three components have influence on the creation the firm value (Δ*Vp*).

**Figure 1**

**The inventory management decision influence on firm value**

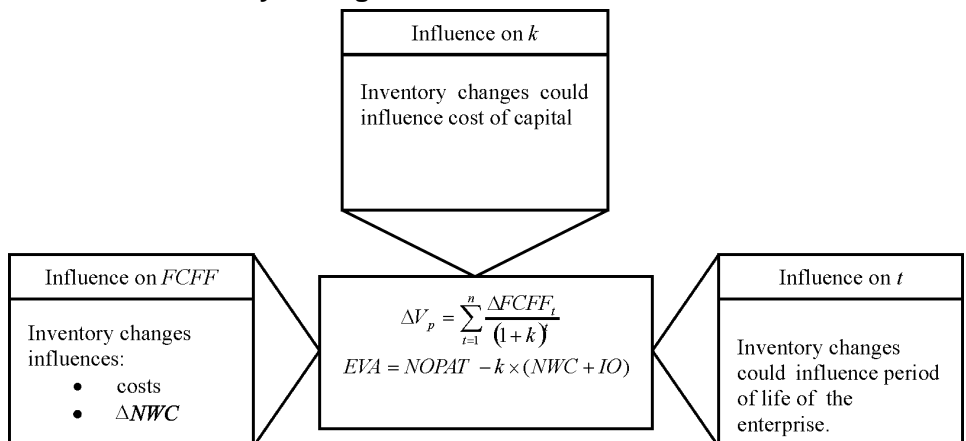

Where: *FCFF* = free cash flows to firm; Δ*NWC* = net working capital growth; *k* = cost of the capital financing the firm; and *t* = the lifetime of the firm and time to generate single *FCFF*.

*Source: own study*.

*Value-Based Inventory Management*

Inventory changes (resulting from changes in inventory management policy of the firm) affect the net working capital level and the level of operating costs of inventory management in a firm as well. These operating costs are result of storage, insurance, transport, obsolescence, wasting and spoilage of inventory).

## 3. EOQ AND VBEOQ

The Economic Order Quantity Model is a model which maximizes the firm's income trough total inventory cost minimization.

**Figure 2**

**EOQ and VBEOQ model**

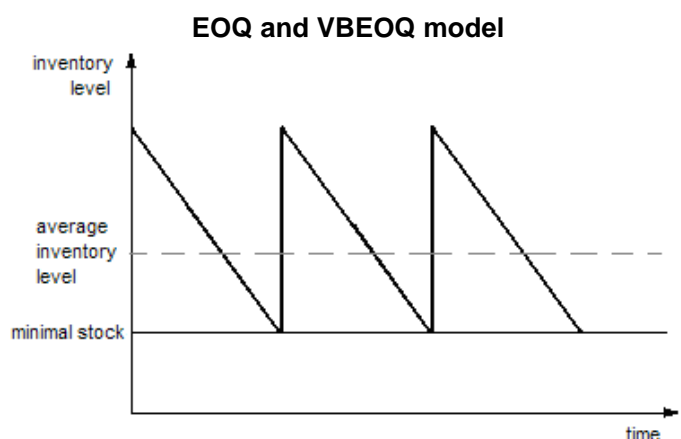

*Source: J. G. Kalberg, K. L. Parkinson, Corporate liquidity: Management and Measurement, IRWIN, Homewood 1993, p. 538.*

The EOQ model requires two equations:

$$EOQ = \sqrt{\frac{2 \times P \times K_z}{C \times v}} = \sqrt{\frac{2 \times P \times K_z}{K_u}} \quad (6)$$

Where: *EOQ* = economic order quantity; *P* = demand for the product/inventory in period (year, month); $K_z$ = cost per order; $K_u$ = holding cost per unit in period (year, month); *C* = holding cost factor; and *v* = purchase cost per unit.

The holding cost factor ($K_u$) is a result of costs[1]:
- Opportunity costs (price of money tied-up in inventory),
- Storage, insurance, transportation, obsolescence, wasting and spoilage costs.

$$TCI = \frac{P}{Q} \times K_z + \left(\frac{Q}{2} + z_b\right) \times v \times C \quad (7)$$

---
[1] *M. Sierpińska, D. Wędzki, Zarządzanie płynnością finansową w przedsiębiorstwie, WN PWN, Warszawa 2002, s. 112.*

Where: *TCI* = total costs of inventory; *Q* = order quantity; and $z_b$ = minimal stock.

**Example 1.** *P* = 220 000 kg; $K_z$ = 31$; *v* = 2$ / 1kg; *C* = 25%. Effective tax rate, *T* = 19%. Cost of capital financing the firm *WACC* = *k* = 15%; $z_b$ = 300 kg.

First we estimate *EOQ*:

$$EOQ = \sqrt{\frac{2 \times 220\,000 \times 31}{0.25 \times 2}} = 5\,223 \text{ kg}.$$

Next we estimate average inventory level:

$$INV_{EOQ=5223} = \frac{5\,223}{2} + 300 = 2\,912 kg \Rightarrow INV_{EOQ=5223} = 2\,912 \times 2 = 5\,824\$$$

$$TCI_{EOQ=5223} = \frac{220\,000}{5\,223} \times 31 + \left(\frac{5\,223}{2} + 300\right) \times 2 \times 0.25 = 2\,762\$.$$

If we chose to order 5 000 kg than the quantity *EOQ* = 5 223 kg, the TCI are:

$$TCI_{Q=5000} = \frac{220\,000}{5\,000} \times 31 + \left(\frac{5\,000}{2} + 300\right) \times 2 \times 0.25 = 2\,764\ \$.$$

We will have greater *TCI*, but if we check how it influences the firm value, we will see that if we decide to order less than *EOQ* suggests, we will increase the firm value:

$$\Delta TCI_{Q=5223 \to Q=5000} = 2\,764 - 2\,762 = 2\ \$,$$

$$INV_{Q=5000} = 2 \times \left(\frac{5\,000}{2} + 300\right) = 5\,600\ \$,$$

$$\Delta INV_{Q=5223 \to Q=5000} = 5\,600 - 5\,824 = -224\ \$,$$

$$\Delta NWC = \Delta INV,$$

$$\Delta V_{Q=5223 \to Q=5000} = 224 - \frac{2 \times (1-0.19)}{0.15} = 213.2\ \$;$$

$$\Delta EVA_{Q=5223 \to Q=5000} = \Delta NOPAT - k \times (\Delta NWC + \Delta OI) = (1-0.19) \times$$
$$\times (-2) - 0.15 \times (-224) = 32\ \$.$$

Because both ΔV and ΔEVA are greater than 0, we see that it will be profitable for the firm to order 5000 kg, not 5223 kg marked by *EOQ*. The *EOQ* model minimizes operational inventory costs, but in firm management we also have opportunity costs of holding inventories. These costs dictate that we will order less than *EOQ* if we want to maximize the firm value. Knowing that we can use *VBEOQ* model:

$$VBEOQ = \sqrt{\frac{2 \times (1-T) \times K_z \times P}{v \times (k + C \times (1-T))}} \qquad (8)$$

Where: *k* = cost of capital financing the firm (*WACC*); and *VBEOQ* = value-based economic order quantity.

*Value-Based Inventory Management*

For Alfa data, we have:

$$VBEOQ = \sqrt{\frac{2\times(1-0{,}19)\times 31\times 220\,000}{2\times(0{,}15+0{,}25\times(1-0{,}19))}} = 3\,959\ kg;$$

$$TCI_{VBEOQ=3959} = \frac{220\,000}{3\,959}\times 31 + \left(\frac{3\,959}{2}+300\right)\times 2\times 0{,}25 = 2\,862\ \$;$$

$$\Delta TCI_{Q=5223\to Q=3959} = 2\,862 - 2\,762 = 100\ \$;$$

$$INV_{VBEOQ=3959} = 2\times\left(\frac{3\,959}{2}+300\right) = 4\,559\ \$;$$

$$\Delta INV_{Q=5223\to Q=3959} = 4\,559 - 5\,824 = -1\,265\ \$;$$

$$\Delta V_{Q=5223\to Q=3959} = 1\,265 - \frac{100\times(1-0{,}19)}{0{,}15} = 725\ \$;$$

$$\Delta EVA_{Q=5223\to Q=3959} = (1-0{,}19)\times(-100) - 0{,}15\times(-1265) = 109\ \$.$$

As we see, both ΔV and ΔEVA are greater than before if the firm order 3 959 kg is marked by VBEOQ. In fact it is the best known possibility.

## 4. POQ AND VBPOQ

Production order quantity model *(POQ)* is the *EOQ* modification which we can use, when we have grater production possibilities than market capacity.

**Figure 3**

**POQ and VBPOQ**

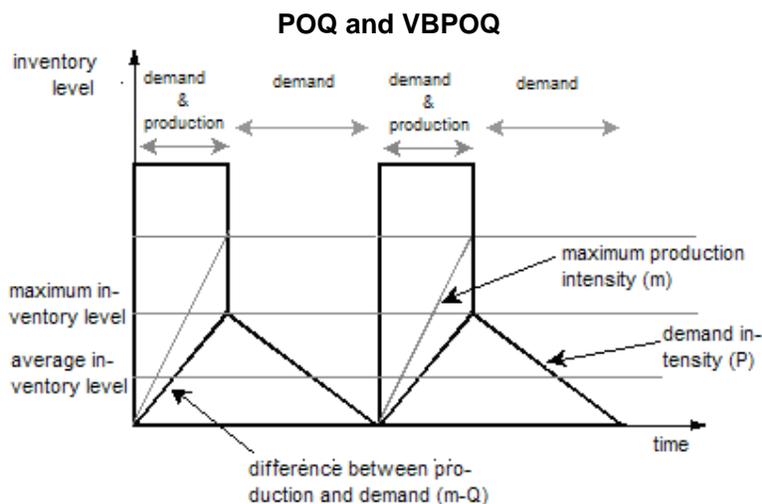

*Source: Z. Sariusz-Wolski, Sterowanie zapasami w przedsiębiorstwie, PWE, Warszawa 2000, p. 162.*

*Institute of Economic Forecasting*

POQ could be estimated as[1]:

$$POQ = \sqrt{\frac{2 \times K_z \times P}{C \times k \times \left(1 - \frac{P}{m}\right)}}, \quad P < m \tag{9}$$

Where: *POQ* = production order quantity; $K_z$ = switch on production cost (setup cost per setup); *P* = demand intensity (how much we can sell annually); *v* = cost per unit; *m* = maximum annual production ability; and *C* = holding cost factor.

$$TCI = \frac{Q}{2} \times \left(1 - \frac{P}{m}\right) \times v \times C + \frac{P}{Q} \times K_z \tag{10}$$

Where: *Q* = production quantity; and *TCI* = total costs of inventories.

$$INV = \frac{Q}{2} \times \left(1 - \frac{P}{m}\right) \tag{11}$$

Where: *INV* = average inventory level.

**Example 2.** Maximum demand, P = 2 500 tons, m = 10 000 tons annually. WACC = k = 15%, C = 25%, T = 19%. $K_z$ = 12 000 $, v = 0,8$.

First we estimate *POQ*:

$$POQ = \sqrt{\frac{2 \times 12\,000 \times 2\,500}{800 \times 0,25 \times \left(1 - \frac{2\,500}{10\,000}\right)}} = 633 \text{ tons.}$$

$$TCI_{POQ=633} = \frac{633}{2} \times \left(1 - \frac{2\,500}{10\,000}\right) \times 800 \times 0,25 + \frac{2\,500}{633} \times 12\,000 = 94\,868\,\$.$$

$$INV_{POQ=633} = \frac{633}{2} \times \left(1 - \frac{2\,500}{10\,000}\right) = 237\,(1000)\,kg \Rightarrow 237 \times 800 = 189\,600\$.$$

Next, we check how firm value will influence the change of production quantity to 90% POQ, 633 × 0,9 = 570 tons:

$$TCI_{POQ=570} = \frac{570}{2} \times \left(1 - \frac{2\,500}{10\,000}\right) \times 800 \times 0,25 + \frac{2\,500}{570} \times 12\,000 = 95\,382\,\$,$$

$$\frac{-\Delta FCFF_{1...\infty}}{0,81} = \Delta TCI_{Q=633 \to Q=570} = 95\,382 - 94\,868 = 514\,\$.$$

$$INV_{POQ=570} = 800 \times INV_{570} = 800 \times \frac{570}{2} \times \left(1 - \frac{2\,500}{10\,000}\right) = 171\,000\,\$,$$

---

[1] *Z. Sariusz-Wolski,* Sterowanie zapasami w przedsiębiorstwie, *PWE, Warszawa 2000, s. 162*

**88**  *Romanian Journal of Economic Forecasting – 1/2008*

*Value-Based Inventory Management*

$$\Delta NWC = (-\Delta FCFF_0) = \Delta ZAP_{Q=633 \to Q=570} = 171\,000 - 189\,600 = (-18\,600)\,\$.$$

$$\Delta V_{Q=633 \to Q=570} = +18\,600 + \frac{-514 \times (1-0,19)}{0,15} = +15\,824\$,$$

$$\Delta EVA_{Q=633 \to Q=570} = (1-0,19) \times (-514) - 0,15 \times (-18600) = 2373,66\$.$$

This shows that, if we will produce less than the quantity POQ, it will create additional value.

VBPOQ can be determined from the following table:

**Table 1**

**VBPOQ**

| Q | TCI | $\Delta$ TCI | INV | $\Delta$ INV | $\Delta$ V | $\Delta$ EVA |
|---|---|---|---|---|---|---|
| 483 | 98337 | 3469 | 144900 | -44700 | 25968 | 3895 |
| 482 | 98391 | 3523 | 144600 | -45000 | 25978 | 3896 |
| 481 | 98445 | 3577 | 144300 | -45300 | 25984 | 3898 |
| 480 | 98500 | 3632 | 144000 | -45600 | 25987 | 3898 |
| 479 | 98555 | 3687 | 143700 | -45900 | 25988 | 3899 |
| 478 | 98612 | 3744 | 143400 | -46200 | 25985 | 3897 |
| 477 | 98668 | 3800 | 143100 | -46500 | 25980 | 3897 |

*Source: Own study.*

We find that VBPOQ gives 479 tons. Table 1 also shows that the costs TCI for VBPOQ will be greater than for POQ, but that VBPOQ ties up less cash in inventories than the POQ what is the source of benefits in lower opportunity costs.

To estimate VBPOQ we also could use the equation:

$$Q_{VBPOQ} = \sqrt{\frac{2 \times P \times K_z \times (1-T)}{v \times \left(1 - \frac{P}{m}\right) \times [k + C \times (1-T)]}}, \quad P < m \quad (12)$$

$$Q_{VBPOQ} = \sqrt{\frac{2 \times 2\,500 \times 12\,000 \times (1-0,19)}{800 \times \left(1 - \frac{2500}{10\,000}\right) \times [0,15 + 0,25 \times (1-0,19)]}} = 479 \text{ tons.}$$

Knowing *VBPOQ*, the firm can better manage inventories and brings the firm closer to realizing the basic financial aim – the firm value maximization.

## 5. Conclusions

Maximization of the owners' wealth is the basic financial goal in enterprise management. Inventory management techniques must contribute to this goal. The modifications to both the value-based *EOQ* model and value-based *POQ* model may be seen in this article. Inventory management decisions are complex. Excess cash tied up in inventory burdens the enterprise with high costs of inventory service and opportunity costs. By contrast, higher inventory stock helps increase income from sales because customers have greater flexibility in making purchasing decisions and the firm decrease risk of unplanned break of production. Although problems connected with optimal economic order quantity and production order quantity remain, we conclude that value-based modifications implied by these two models will help managers make better value-creating decisions in inventory management.